# THEORETICAL INVESTIGATION OF COHERENT SYNCHROTRON RADIATION INDUCED MICROBUNCHING INSTABILITY IN TRANSPORT AND RECIRCULATION ARCS*

C.-Y. Tsai[#], Department of Physics, Virginia Tech, Blacksburg, VA 24061, USA
D. Douglas, R. Li, and C. Tennant, Jefferson Lab, Newport News, VA 23606, USA

*Abstract*

The coherent synchrotron radiation (CSR) of a high brightness electron beam traversing a series of dipoles, such as recirculation or transport arcs, may lead to the microbunching instability. We extend and develop a semi-analytical approach of the CSR-induced microbunching instability for a general lattice, based on the previous formulation with 1-D CSR model [1] and apply it to investigate the physical processes of microbunching amplification for two example transport arc lattices. We find that the microbunching instability in transport arcs has a distinguishing feature of multistage amplification (e.g, up to 6[th] stage for our example arcs in contrast to two stage amplification for a 3-dipole chicane [2]). By further extending the concept of stage gain as proposed by Huang and Kim [2], we developed a method to quantitatively characterize the microbunching amplification in terms of iterative or staged orders that allows the comparison of optics impacts on microbunching gain for different lattices. The parametric dependencies and Landau damping for our example lattices are also studied. Excellent agreement of the gain functions and spectra from Vlasov analysis with results from ELEGANT is achieved which helps to validate our analyses.

## INTRODUCTION

As is well known, CSR effects have been one of the most challenging issues associated with the design of magnetic bunch compressor chicanes for X-ray FELs and linear colliders. Typically, CSR is emitted not only for a wavelength range comparable or longer than the bunch length duration but also for shorter wavelengths if the bunch charge density is modulated at such wavelength range (so called microbunching). Such coherent radiation effects, which have been confirmed both in numerical simulation and experiments, can result in undesirable beam quality degradation. The aforementioned works were, however, mostly focused on studies of magnetic bunch compressor chicanes. Recently the superconducting radio frequency recirculation linac facilities (e.g., free-electron laser facility or electron cooler for collider machine) have been brought to attention [3-5]. The recirculation or transport arcs are necessary elements in such facilities. However, similar studies on such systems with multiple modular dipoles as in a transport or recirculation arc were mostly focused on transverse beam dynamics while longitudinal phase space degradation has very limited discussions. In this paper, we pay attention to the CSR effects on the longitudinal beam dynamics as beam traversing around the recirculation arcs, particularly on the CSR-induced microbunching issues. The reasons to be concerned about the microbunching instability are based on the two facts: first, such system is characteristic of long transport lines with many dipole magnets whether in a single or multiple pass operation; the other, the system typically delivers the beam which features high beam quality, i.e. high brightness, for the use in next-generation light sources or hadron cooling in high-energy electron recirculating cooling rings. Thus, the seeds that cause density modulation, derived from either the density ripples from upstream injector or longitudinal space charge especially for low energy beam, would be extremely possible to lead to microbunching instability. Therefore further investigation of CSR-induced microbunching effects and its detailed physical processes are of critical importance and may shed light on how to improve designs for future lattices.

## METHODS

The CSR-induced microbunching instability can be formulated with the linearized Vlasov equation in typical bunch compressors [1]. This model makes coasting-beam approximation and assumes steady-state 1-D CSR [6] with negligible shielding (boundary) effects. By the method of characteristics, the linearized Vlasov equation can be rewritten as the general form of Volterra integral equation below [1]:

$$g_k(s) = g_k^{(0)}(s) + \int_0^s K(s,s')g_k(s')ds'$$
$$K(s,s') = \frac{ikr_e n_b}{\gamma} C(s)C(s')R_{56}(s' \to s)Z(kC(s'),s')$$
$$\times [\text{Landau damping}]$$
(1)

where the [Landau damping] term can be expressed as

$$\exp\left\{\frac{-k^2}{2}\left[\varepsilon_{x0}\left(\beta_{x0}R_{51}^2(s,s') + \frac{R_{52}^2(s,s')}{\beta_{x0}}\right) + \sigma_\delta^2 R_{56}^2(s,s')\right]\right\}$$
(2)

Here the kernel function $K(s,s')$ describes the CSR interaction, $g_k(s)$ the resultant bunching factor as a function of the longitudinal position given the wavenumber $k$, and $g_k^{(0)}(s)$ is the bunching factor without CSR.

---

In this work, we have extended the formulation to a general transport line by input of the transport functions of a lattice from a common particle tracking code, e.g. ELEGANT [7] and developed a program to solve Eq. (1) self-consistently or Eq. (3) (below) iteratively [8]. For the latter approach, we adapt, motivated by [2], the above integral equation (Eq. (1)) by iteratively finding the staged solutions. The advantage of using this method is to facilitate us exploring up to which stage the overall CSR process can be described, by direct comparison with the self-consistent solutions. The staged iterative solution for *n*-th order can be defined as

$$\mathbf{g}_k^{(n)} = \left(\sum_{m=0}^{n} \mathbf{K}^m\right) \mathbf{g}_k^{(0)} \quad (3)$$

where we have already expressed the bunching factors in vector form for discrete s values and the kernel $K(s,s')$ in matrix form, denoted as **K**. It can be easily shown that Eq. (1) and Eq. (3) are equivalent, i.e. $g_k(s)$ is a component of $\mathbf{g}_k^{(n)}$ at some specific location when $n \to \infty$ provided the sum converges. With a given wavenumber $k$, the CSR-induced microbunching gain is then defined, based on Eq. (1), as $G(s) = |g_k(s)/g_k^{(0)}(0)|$. Note that $g_k$ is in general a complex quantity, or a phasor in complex domain. Setting $\tilde{G}^{(n)}(s) = \mathbf{g}_k^{(n)}(s)/\mathbf{g}_k^{(0)}(0)$, we define the stage gain as the amplitude of the phasor $\tilde{G}^{(n)}(s)$, i.e. $G^{(n)}(s) = |\tilde{G}^{(n)}(s)|$.

To compare the CSR gains contributed from individual stages, we further define the final stage gain phasor up to a certain order $M$ as

$$\tilde{G}_f^{(M)} = \tilde{G}^{(M)}(s = s_f) = \tilde{G}_0 + \tilde{G}_1 I_b + ... + \tilde{G}_M I_b^M = \sum_{m=0}^{M} \tilde{G}_m I_b^m$$

In order to extract the net effect caused by lattice optics, the above expression can be further formulated as

$$\tilde{G}_f^{(M)} = \sum_{m=0}^{M} A^m d_m^{(\lambda)} \left(\frac{I_b}{\gamma}\right)^m \quad (4)$$

where $A = -0.94 + 1.63i$, $I_b$ the beam current, $\gamma$ the relativistic factor of beam energy, and $d_m^{(\lambda)}$ (given some modulation wavelength $\lambda = 2\pi/k$) now reflects the effects from lattice optics at *m*-th stage and Landau damping through beam emittance and energy spread. Again we remind the connection between Eq. (3) and Eq. (4) $G_f^{(M)} = |\tilde{G}_f^{(M)}| = G^{(M)}(s = s_f)$. As mentioned, we implement the above semi-analytical methods (self-consistent and iterative approach) by input of the transport functions from ELEGANT [7], and then solve Eq. (1) and/or Eq. (3) for the gain function $G(s)$ or gain spectrum $G_f(\lambda)$.

## SIMULATION RESULTS

### CSR Microbunching Gain Analysis

To demonstrate how our new methods can be used to calculate of microbunching gain spectrum for general arc lattices, we take two 180° transport arc lattices as our study examples. For the detailed description of the two example lattices, we refer the interested reader to [9]. The first example lattice, a 1.3 GeV recirculation arc with large momentum compaction function (Example 1), is a second order achromat and globally isochronous with a large dispersion modulation across the entire arc. This allows an assessment of the impact of large momentum compaction oscillation, large dispersion, and the absence of multiple periodic isochronicity on CSR microbunching gain. In contrast to the first example, the second example, a 1.3 GeV recirculation arc with small momentum compaction function (Example 2), is also a second order achromat but designed to be a locally isochronous lattice within superperiods which insure that the bunch length is the same at phase homologous CSR emission sites. The small dispersion and its slope result not only in a small momentum compaction in each superperiod but the modulation of the momentum compaction through the system is also small; thus potentially providing some limitation on CSR gain.

Table 1: Initial Beam and Twiss Parameters for the Two Example Arc Lattices

| Name | Example 1 (large $R_{56}$) | Example 2 (small $R_{56}$) | Unit |
|---|---|---|---|
| Beam energy | 1.3 | 1.3 | GeV |
| Bunch current | 65.5 | 65.5 | A |
| Normalized emittance | 0.3 | 0.3 | μm |
| Initial beta function | 35.81 | 65.0 | m |
| Initial alpha function | 0 | 0 | |
| Energy spread (uncorrelated) | $1.23 \times 10^{-5}$ | $1.23 \times 10^{-5}$ | |

Table 1 summarizes the initial beam and Twiss parameters of the two high-energy recirculation arcs based on [9]. The CSR-induced microbunching gains for the two recirculation arcs are shown in Figs. 1 and 2. The two upper figures demonstrate the evolution of the self-consistent gain as a function of *s* for three different modulation wavelengths. One can see in Fig. 1 the shorter wavelengths enhance the Landau damping through Eq. (2) while longer wavelengths feature negligible CSR effect [6]. The two bottom figures show the gain spectra $G_f(\lambda)$ at the exits of the lattices as a function of modulation wavelength, from which one can obviously see the difference between them: Example 1 is vulnerable to CSR effect while Example 2 is not. To validate our semi-analytical approach, we also simulate the two example lattices by ELEGANT [7]. Extensive convergence studies were performed for ELEGANT tracking [10], and the gain spectra from ELEGANT show good agreement with our semi-analytical results (see Figs. 1 and 2). Because of very high gain for Example 1 lattice, we notice that ELEGANT simulations encounter a big challenge for microbunching gain calculation based on particle tracking algorithm, so special cares are required [10]. From the iterative approach in Eq. (3), we also find an interesting feature of microbunching in the two example arcs: in contrast to the 2-stage amplification exhibited in a conventional bunch compressor chicane [2], the CSR-induced microbunching gain in the recirculating arcs (each consisting of 24 dipoles) requires up to 6$^{th}$ stage amplification, as shown in bottom figures of Figs. 1 and 2.

To further examine the feature of multistage CSR gain for such systems, we extract the coefficients $d_m^{(\lambda)}$ (here $\lambda$ is chosen that corresponds to the maximal gain; for Example 1 and 2 lattices, $\lambda_{opt}$ = 36.82 and 19 μm, respectively) in Eq. (4) by fitting each staged solutions obtained from Eq. (3) so that we can quantify and compare the individual stage gains with inclusion of lattice optics effects and beam parameters $\varepsilon_{x0}$ and $\sigma_\delta$ associated with Landau damping. Figure 3 shows the bar charts representing the final gains as functions of beam current and stage index for both arcs. Since $d_m^{(\lambda)}$ are independent of beam current and energy, they can be used to obtain a current scaling law of CSR gain for each given wavelength. Figure 4 compares the final overall gain from Eq. (4) and Eq. (1) for different currents in the case of Example 1 lattice, at a selected wavelength that is in the vicinity of optimal wavelength for maximal gain.

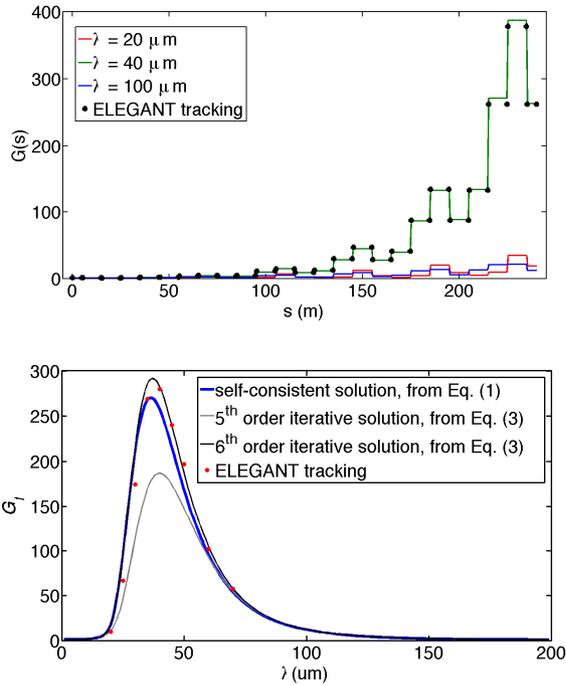

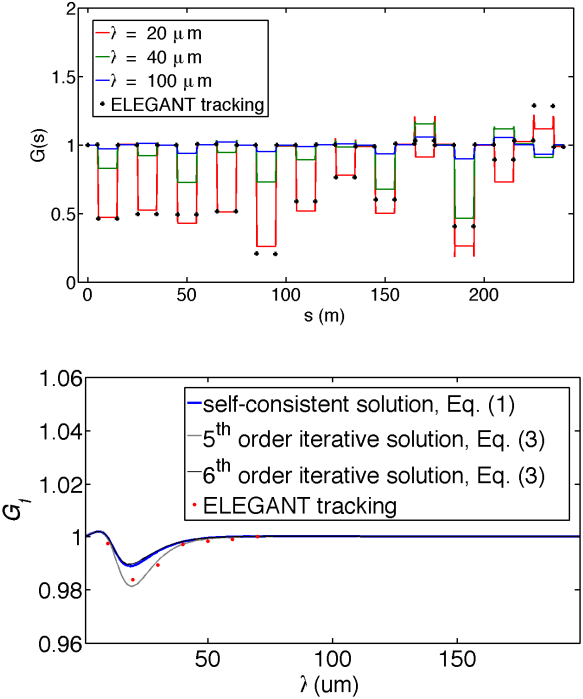

Figure 2: (top) CSR gain functions $G(s)$, where dots are for $\lambda$ = 20 μm; (bottom) gain spectrum $G_f(\lambda)$ as a function of initial modulation wavelength for Example 2 lattice.

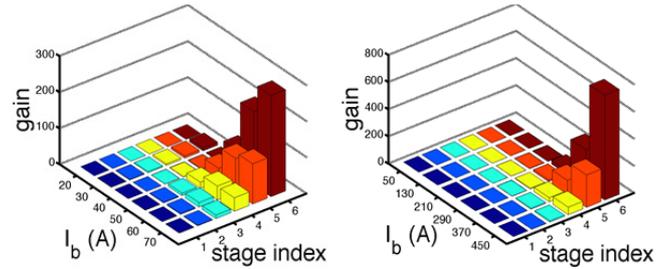

Figure 3: Bar chart representation of staged gains (at the exit of lattice) for several different beam currents for the two example arcs, (left) Example 1 lattice; (right) Example 2 lattice, where for Example 1 and 2 lattices, $\lambda_{opt}$ = 36.82 and 19 μm, respectively. Note the current scale ($I_b$) and vertical scale are different.

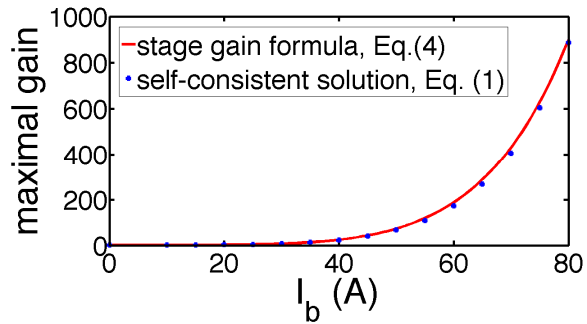

Figure 1: (top) CSR gain functions $G(s)$, where dots are for $\lambda$ = 40 μm; (bottom) gain spectrum $G_f(\lambda)$ as a function of initial modulation wavelength for Example 1 lattice.

Figure 4: The current dependence of maximal CSR gain for the Example 1 lattice. Solid red line from Eq. (4) with $M$ = 6, and blue dots from Eq. (1).

## Parametric Dependencies and Landau Damping

For a gain spectral curve, we are most interested in the maximal gain and the corresponding *optimum* wavelength. Figure 5 shows the parametric dependencies of maximal gains and optimum wavelengths as functions of energy spread (for zero beam emittance) and beam emittance (for zero beam energy spread) for Example 1 lattice. We can see that both larger energy spread and beam emittance result in more Landau damping and the optimum wavelength drifts toward longer wavelengths to compensate the overall damping. It is also found in the two top figures the optimum wavelengths scale linearly and radically for beam energy spread and emittance, respectively. In the two bottom figures the maximal gains behave proportionally to $\sigma_\delta^{-8}$ and $\varepsilon_{x0}^{-4}$.

After carefully examining the parametric dependencies for Example 2 lattice and combining the results of the two lattices [8], we find the behaviour of optimum wavelength $\lambda_{opt}$ features a linear relationship to beam energy spread $\sigma_\delta$ (for zero beam emittance) as

$$\lambda_{opt} \propto \sigma_\delta \quad (5)$$

Here we note that Ref. [11] gives a linear expression on how the optimum wavelength relates to beam energy spread assuming 2-stage amplification in a typical bunch compressor for zero-emittance beam. Although our example lattices feature 6$^{th}$ stage amplification, the optimum wavelength is also characteristic of linear relation to beam energy spread (Eq. (5)). Whether such observations reflect the general behaviour for multistage amplification and/or arbitrary lattice is still a question for further investigation. Similarly, we obtain a relation for the dependence of the optimum wavelength on beam emittance (for zero energy spread beam) as

$$\lambda_{opt} \propto \sqrt{\varepsilon_{x0}} \quad (6)$$

Furthermore, we simulate and analyse several lattices with varying amplification stages [8] and find the maximal gain can be scaled in the case of zero beam emittance at *M*-th stage as,

$$G_{f,\max} \propto \frac{1}{\sigma_\delta^{(4/3)M}} \quad (7)$$

and in the case of zero energy spread at *M*-th stage as,

$$G_{f,\max} \propto \frac{1}{\varepsilon_{x0}^{(2/3)M}} \quad (8)$$

For the two Example lattices, $M \approx 6$. The observation of such scaling (Eq. (5) and Eq. (6)) also corresponds well to the gain behaviour in a typical bunch compressor chicane featuring two-stage amplification (*M* = 2) [10].

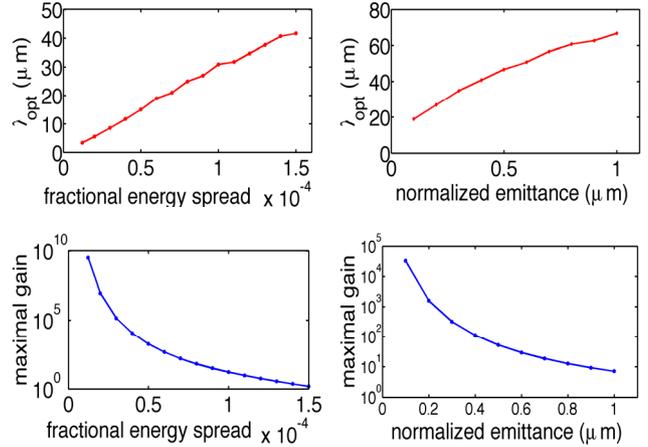

Figure 5: Optimum wavelengths (upper) and maximal CSR gains (lower) as functions of beam energy spread with zero emittance and finite beam emittance with vanishing energy spread for Example 1 lattice.

To further illustrate how the multi-stage amplification contributes to and how the lattice optics impacts on the microbunching development, we create in Fig. 6 the "quilt" patterns of $R_{56}(s' \to s)$ (shown in the kernel of Eq. (1)) behaviour for the two example lattices in order to clearly identify the enhancement or suppression of microbunching along the beamline by lattice optics. For a planar and uncoupled lattice, the term $R_{56}(s' \to s) = R_{56}(s) - R_{56}(s') + R_{51}(s')R_{52}(s) - R_{51}(s)R_{52}(s')$. The upper left area in the figures has vanishing value due to causality. It is obvious that in the left figure the block areas with large amplitude of $R_{56}(s' \to s)$, particularly the bottom right deep red blocks, exist and can accumulate the CSR gain. To be specific, for Example 1 lattice (left figure), energy modulation at $s' \approx$ 15 m will cause density modulation at $s \approx$ 60 m, that causes CSR-induced energy modulation at this location. Then such modulation propagate by $R_{56}(s' \to s)$ from $s' \approx$ 60 m to $s \approx$ 100 m, and so on. It is this situation that causes multi-stage CSR amplification. Here we note that more complete analysis needs to take Landau damping factor into account (Eq. (2)) [8]. In contrast, the situation in the right figure for Example 2 lattice is more alleviated because of the smaller amplitudes of $R_{56}(s' \to s)$.

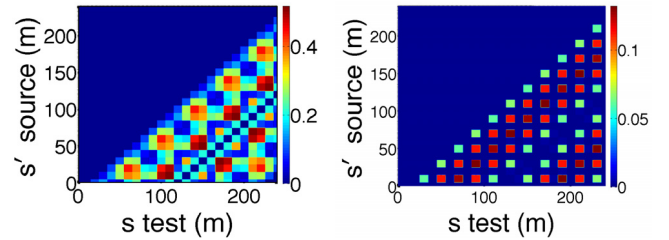

Figure 6: $R_{56}(s' \to s)$ "quilt" patterns for the two Example lattice arcs: Example 1 (left) and Example 2 (right).

## SUMMARY

In this paper we extended previous work [1, 2] to study the CSR-induced microbunching gain in transport and recirculation arcs, by finding both the self-consistent [Eq. (1)] and iterative [Eq. (3)] solutions. Different from that of typical bunch compressor chicane, CSR gain is characteristic of multistage amplifications, e.g. up to 6-th stage in our two example lattices (Figs. 1 and 2). Then we quantified and compared the CSR gains contributed from individual stages for the two lattices with exclusion of beam current and energy dependence (Fig. 3) and demonstrate a current scaling law for close-to-maximum CSR gains for a fixed wavelength in the vicinity of optimal one (Fig. 4).

The parametric dependencies of optimum wavelength and maximal gain on beam energy spread and beam emittance are summarized in Eq. (5-8) from our semi-analytical simulations for Example 1 lattice. We also note that whether such semi-analytical observations reflect the general behavior for multistage amplification and/or arbitrary lattice is still a question for further investigation.

We finally presented in Fig. 6 the $R_{56}(s' \to s)$ impact on the overall CSR gain for both example lattices and found that for Example 1 lattice there exist several cumulated large-amplitude areas so that the gain eventually builds up (i.e. multi-stage amplification) while for Example 2 this situation is well controlled at the same range of beam current scale. As shown in Fig. 3, with the same current range, Example 1 can reach much larger gain than Example 2. This demonstrates the impact of lattice design on the control of microbunching.

## ACKNOWLEDGMENT

This work is supported by Jefferson Science Associates, LLC under U.S. DOE Contract No. DE-AC05-06OR23177.

## REFERENCES

[1] S. Heifets *et al.*, Coherent synchrotron radiation instability in a bunch compressor, Phys. Rev. ST Accel. Beams **5**, 064401 (2002)

[2] Z. Huang and K. -J. Kim, Formulas for coherent synchrotron radiation microbunching in a bunch compressor chicane, Phys. Rev. ST Accel. Beams **5**, 074401 (2002)

[3] R. C. York, 5 upgradable to 25 keV free electron laser facility, Phys. Rev. ST Accel. Beams **17**, 010705 (2014)

[4] Y. Zhang and J. Bisognano Eds., Scientific requirements and conceptual design for a polarized medium energy electron-ion collider at Jefferson Lab, arXiv: 1209.0757v2 [physics.acc-ph]

[5] P. Piot *et al.*, Longitudinal phase space manipulation in energy recovery linac-driven free-electron lasers, Phys. Rev. ST Accel. Beams **6**, 030702 (2003)

[6] Ya. S. Derbenev *et al.*, Microbunch radiative tail-head interaction, TESLA-FEL-Report 1995-05

[7] M. Borland, elegant: A Flexible SDDS-Compliant Code for Accelerator Simulation, Advanced Photon Source LS-287 (2000)

[8] C. -Y. Tsai *et al.*, to be published

[9] D. R. Douglas *et al.*, Control of coherent synchrotron radiation and microbunching effects during transport of high brightness electron beams, arXiv: 1403.2318v1 [physics.acc-ph]

[10] C.-Y. Tsai and R. Li, Simulation of coherent synchrotron radiation induced microbunching gain using ELEGANT, JLAB-TN-14-016

[11] E. L. Saldin *et al.*, Klystron instability of a relativistic electron beam in a bunch compressor, Nucl. Instrum. Methods Phys. Res. Sect. A 490, 1 (2002)